\documentclass[prl,twocolumn,amssymb,floatfix]{revtex4-1}
\usepackage{amsmath} 
\usepackage{amsfonts} 
\usepackage{amssymb}
\usepackage{bbm}
\usepackage{epsfig}
\usepackage{graphics}
\usepackage{graphicx}

\def\ptps{{Prog.\ Theor.\ Phys.\ Suppl. \ }}

\newcommand{\lsim}{\,\lower2truept\hbox{${<\atop\hbox{\raise4truept\hbox{$\sim$}}}$}\,}
\newcommand{\gsim}{\,\lower2truept\hbox{${>\atop\hbox{\raise4truept\hbox{$\sim$}}}$}\,}
\newcommand{\pp}{~~~.}
\newcommand{\vv}{~~~,}

\newcommand{\be}{\begin{equation}}
\newcommand{\ee}{\end{equation}}
\newcommand{\bea}{\begin{eqnarray}}
\newcommand{\eea}{\end{eqnarray}}
\newcommand{\beann}{\begin{eqnarray*}}
\newcommand{\eeann}{\end{eqnarray*}}
\newcommand{\nn}{\nonumber}

\newcommand{\divvec}[1]{\mathbf{\nabla\hspace{-.3em}\cdot\hspace{-.2em} #1}}
\newcommand{\scalprod}[2]{\mathbf{#1 \hspace{-.2em}\cdot\hspace{-.2em} #2}}

\begin{document}

\title{Very large scale structures in growing neutrino quintessence}

\author{N. Wintergerst$^1$, V. Pettorino$^{1,2}$, D. F. Mota$^3$, C. Wetterich$^1$}
\affiliation{$^1$ Institut f\"ur Theoretische Physik, Universit\"at Heidelberg, Philosophenweg 16, D-69120 Heidelberg, Germany,\\
$^2$ Italian Academy for Advanced Studies in America, Columbia University, 1161 Amsterdam Avenue, New York, NY 10027, USA,\\
$^3$ Institute of Theoretical Astrophysics, University of Oslo, 0315 Oslo, Norway}

\begin{abstract}
  A quintessence scalar field or cosmon interacting with neutrinos can have important effects on cosmological structure formation. Within growing neutrino models the coupling becomes effective only in recent times, when neutrinos become nonrelativistic, stopping the evolution of the cosmon. This can explain why dark energy dominates the Universe only in a rather recent epoch by relating the present dark energy density to the small mass of neutrinos. Such models predict the presence of stable neutrino lumps at supercluster scales ($\sim 200$ Mpc and bigger), caused by an attractive force between neutrinos which is stronger than gravity and mediated by the cosmon. We present a method to follow the initial nonlinear formation of neutrino lumps in physical space, by integrating numerically on a 3D grid nonlinear evolution equations, until virialization naturally occurs. As a first application, we show results for cosmologies with final large neutrino average mass $\sim 2$ eV: in this case, neutrino lumps indeed form and mimic very large cold dark matter structures, with a typical gravitational potential $10^{-5}$ for a lump size $\sim 10$ Mpc, and reaching larger values for lumps of about $200$ Mpc. A rough estimate of the cosmological gravitational potential at small $k$ in the nonlinear regime, $\Phi_{\nu} = 10^{-6} (k/k_0)^{-2}\,,\,\, 1.2\cdot10^{-2}$ h/Mpc $< k_0 < 7.8\cdot10^{-2}$ h/Mpc, turns out to be many orders of magnitude smaller than an extrapolation of the linear evolution of density fluctuations. The size of the neutrino-induced gravitational potential could modify the spectrum of CMB anisotropies for small angular momenta.

\end{abstract}

\maketitle

\section{Introduction}

The presence of an interaction between a quintessence scalar field or cosmon and other species in the Universe \cite{wetterich_1995} \cite{amendola_2000} influences the nature and properties of dark energy, with relevant effects on structure formation \cite{perrotta_baccigalupi_2002, nusser_etal_2005, koivisto_2005, kesden_kamionkowski_2006,mainini_bonometto_2006, farrar_rosen_2007, bean_etal_2007, cabral_etal_2009, keselman_etal_2009, afshordi_etal_2005, lavacca_etal_2009,bernardini_etal_2009}. Recently, {\emph N}-body simulations have been performed for dark matter particles interacting with dark energy \cite{baldi_etal_2008, maccio_etal_2004, li_zhao_2009}. We concentrate here on growing neutrino models \cite{amendola_etal_2007, wetterich_2007}, where the neutrino-cosmon coupling can explain the ``why now?'' problem of dark energy. It has been recently shown \cite{mota_etal_2008} that this predicts the existence of very large (supercluster) structures.

Indeed, the key ingredient of growing neutrino quintessence is the pre\-sence of a coupling between dark energy and neutrinos. The latter have a mass that grows with the evolution of the Universe - $m_\nu(\phi)$ being a function of the cosmon field $\phi$. The cosmon-neutrino coupling $\beta$ is given by the logarithmic derivative $\beta = -d\ln{m_\nu}/d \phi$. Since neutrino masses become cosmologically relevant only for redshift $z \approx 5$, this framework can naturally answer the question why only in recent times dark energy leads the Universe expansion to accelerate, thus providing a solution to the coincidence problem. 

In these models, as long as neutrinos stay relativistic, the coupling plays no role and dark energy tracks the background, along the attractor trajectories characterizing the cosmon evolution in the presence of an exponential potential \cite{wetterich_1995, wetterich_1988, ratra_peebles_1988, ferreira_joyce_1998, copeland_etal_1998}. When neutrinos become nonrelativistic, the coupling between neutrinos and quintessence becomes relevant and almost stops the evolution of the cosmon. Then dark energy starts to resemble a cosmological constant with roughly the value of the exponential potential at the end of the attractor era. The transition from the attractor solution to an almost static solution is therefore strictly connected to a cosmological event, that is neutrinos becoming nonrelativistic. This ``trigger event'' leads dark energy to dominate over cold dark matter, naturally starting the recent era of accelerated expansion.

Growing neutrino quintessence requires a cosmon-neutrino interaction somewhat stronger than gravity - typically the attraction between nonrelativistic neutrinos exceeds gravity by a factor $10^{3}$. This coupling is substantially larger than a possible cosmon coupling to atoms. This may be motivated by the particular particle physics mechanism responsible for the neutrino mass, which typically involves a heavy singlet field and not only the standard Higgs doublet \cite{wetterich_2007}. Even much larger neutrino couplings leading to a strongly coupled ``acceleron-neutrino fluid'' have been investigated within mass varying neutrino
models \cite{fardon_etal_2004, afshordi_etal_2005, bjaelde_etal_2007}. In particular, mass varying neuitrino models employ a scalar field with a mass much larger than the Hubble parameter. For growing neutrino models, in contrast, the time dependent cosmon mass equals the Hubble parameter up to a factor of order one, similar to many models of coupled quintessence. (For coupled quintessence with neutrinos at the linear level see ref. \cite{bean_etal_2007, mota_etal_2008, brookfield_etal_2005, ichiki_keum_2007, franca_etal_2009}.) For this reason, the cosmon and the neutrinos always have to be treated as separate ingredients rather than as a common fluid. We remark that the small time dependent cosmon mass follows naturally from possible explanations of an exponential potential in forms of an asymptotically vanishing dilatation anomaly \cite{wetterich_2008}. It requires no additional fine tuning of particle physics parameters. 

Furthermore, in growing neutrino cosmologies the coupling is ineffective for most of the cosmological evolution and only becomes active when neutrinos become nonrelativistic, relating naturally dark energy and neutrino properties. 
In view of bounds on the present neutrino mass, $m_\nu(t_0) < 2.3 {\rm eV}$ \cite{amsler_etal_2008}, and the time
dependence of $m_\nu$, which makes the mass even smaller in the past, the time
when neutrinos become nonrelativistic is typically in the recent history of
the Universe, say $z_{{\rm NR}} \approx 5-10$ \cite{amendola_etal_2007}. It is only from this time on that neutrinos start feeling the effects of the coupling, manifesting effectively as a new attractive interaction between neutrinos. The `fifth force' responsible for the formation of neutrino lumps ``switches on'' only in rather recent cosmology. 

Neutrino fluctuations on length scales larger than the free streaming length are still present at
$z_{{\rm NR}}$, and they start growing for $z < z_{{\rm NR}}$ with a large growth rate. As illustrated in \cite{mota_etal_2008}, this
opens the possibility that neutrino perturbations rapidly grow nonlinear on supercluster scales and beyond. Neutrino nonlinear fluctuations later turn into bound neutrino lumps of the type discussed in
\cite{brouzakis_etal_2007}, thus opening a window for observable effects of the growing neutrino scenario. The linear analysis \cite{mota_etal_2008} has already provided an estimate of this effect as a function of redshift and scale. Typically large scale neutrino fluctuations with size $\gsim 10$ Mpc become nonlinear at a redshift $z \approx 1$.

As noted in \cite{mota_etal_2008}, a continuation of the linear evolution beyond the time when the neutrino fluctuations are of order unity can easily produce erroneous results. In the linear approximation, the neutrino density contrast would quickly reach huge values, producing a very large gravitational potential. Then a very strong ISW effect would seemingly indicate a strong conflict with the cosmic microwave anisotropies (cf. ref. \cite{franca_etal_2009} for a linear analysis). The true physics differs very strongly from the linear behavior, for example by the asymmetry between very large positive density contrasts, while a negative density contrast is bounded by $\delta_{\nu} \geq -1$ since the neutrino density cannot be negative. In order to understand the true gravitational potentials that will be generated by the neutrino lumps, one has to understand the nonlinear dynamics of how local neutrino lumps form, how they are distributed in mass and size and how they may merge into larger lumps as time goes on.

In this work we investigate the formation of individual neutrino lumps at a nonlinear level and in the Newtonian limit. One may wonder if a spherical collapse approach suffices to give a meaningful description of the lumps. We find that this is not the case, as the major force driving neutrinos to collapse is not gravity but the additional fifth force introduced by the coupling to the cosmon and dominant once neutrinos decouple from the background expansion. The evolution of the effective cosmic scale factor for the space occupied by the neutrino lump is only a subleading effect, in contrast to the formation of dark matter halos. We have therefore developed a numerical method for solving the hydrodynamic equations for the neutrino fluid, coupled to the cosmon and gravity. We have also included dark matter, but this is a subleading effect.

The nonlinear analysis developed in this work provides a self-consistent way of analyzing the growth of neutrino perturbations in growing neutrino models. We show that neutrinos indeed form stable structures on large sub-horizon scales and we estimate the properties of the lump as a function of redshift and scale. In particular, we compute the gravitational potential of the lump at a time when the collapse ends due to virialization. This is a key quantity for an estimate of the effects of neutrino lumps on large scale structure - as large scale peculiar velocities - or on the cosmic microwave anisotropies in form of the integrated Sachs - Wolfe (ISW) effect.

This paper is organized as follows. In section \ref{gnm} we recall the framework of growing neutrino cosmologies in which neutrino lumps form. In section \ref{nl} we introduce the set of equations describing the evolution of neutrino overdensities at a nonlinear level. We comment on the methods used in the numerical integration and present our results for the case of large present neutrino masses in section \ref{sec:results}. Section \ref{sec:ini_conds} discusses the initial conditions used for the nonlinear analysis. In section \ref{sec:rel_nonrel} we relate the nonlinear equations to relativistic linear ones. Finally we draw our conclusions in section \ref{conclusions}.
 
\section{Growing neutrino cosmologies} \label{gnm}

Growing neutrino models are described by the set of equations illustrated in \cite{mota_etal_2008} both for the evolution of the homogeneous and isotropic background and for linear perturbations. Here we recall for convenience the essential ingredients characterizing these models.
At the background level, the Universe evolves in time according to the Friedmann and acceleration equations:
\be \label{f1} {\cal H}^2 \equiv \left(\frac{a'}{a}\right)^2 = \frac{a^2}{3} \sum_\alpha \rho_\alpha - \frac{k}{a^2} \ee
and
\be \label{f2} \frac{a''}{a} = {\cal H}^2 - \frac{a^2}{6} \sum_\alpha \left[\rho_\alpha(1 + 3 w_\alpha)\right] \vv \ee
where primes denote derivatives with respect to conformal time $\tau$, the sum is taken over all components $\alpha$ of the energy density in the Universe. We use $k = 0$ for a spatially flat background. The equation of state $w_\alpha$ is related to the energy density $\rho_\alpha$ for each species in the usual way, $w_\alpha \equiv p_\alpha/\rho_\alpha$.
A crucial ingredient in this model is the dependence of the neutrino mass on
the cosmon field $\phi$, as encoded in the dimensionless cosmon-neutrino
coupling $\beta$, \be \beta \equiv - \frac{d \ln{m_\nu}}{d \phi} \pp
\ee For increasing $\phi$ and $\beta < 0$ the neutrino mass increases with
time \be m_{\nu} = \bar{m}_{\nu} e^{-{{\beta}} \phi} \vv \ee where
$\bar{m}_{\nu}$ is a constant. The coupling $\beta$ is chosen here to be a constant but can be, in general, a function of $\phi$, as
proposed in \cite{wetterich_2007} within a particle physics model, leading to similar effects. The cosmon field $\phi$ is normalized in units of the reduced Planck mass $M = (8 \pi G_N)^{-1/2}$, and $\beta^2$ gives the strength of the cosmon mediated interaction. The case $\beta \sim 1$ corresponds to a strength comparable to gravity. For a given
cosmological model with a given time dependence of $\phi$, one can determine
the time dependence of the neutrino mass $m_\nu(t)$. For three degenerate
neutrinos the present value of the neutrino mass $m_\nu(t_0)$ can be related
to the energy fraction in neutrinos ($h \approx 0.72$) \be \Omega_{\nu} (t_0) = \frac{3 m_\nu
  (t_0)}{94\, {\rm eV} h^2} \,\, . \ee

The dynamics of the cosmon can be inferred from the Klein
Gordon equation, now including an extra source due to the neutrino coupling,

\be \label{kg} \phi'' + 2{\cal H} \phi' + a^2 \frac{dU}{d \phi} = a^2 \beta
(\rho_{\nu}-3 p_{\nu}) \,\, , \ee with $\rho_\nu$ and $p_\nu = w_\nu \rho_\nu$ 
the energy density and pressure of the neutrinos. We choose
an exponential potential \cite{wetterich_1988, ratra_peebles_1988, ferreira_joyce_1998, barreiro_etal_2000, wetterich_2007}:

\be \label{pot_def} V(\phi) = M^2 U(\phi) = M^4 e^{- \alpha \phi} \vv \ee
where the constant $\alpha$ is one of the free parameters of our model and determines the amount of nonnegligible dark energy at early times. Current bounds constrain it to be of the order $\alpha \sim 10$ or bigger \cite{doran_etal_2007}. 

The homogeneous energy density and pressure of the scalar field $\phi$ are defined
in the usual way as \be \label{phi_bkg} \rho_{\phi} = \frac{\phi'^2}{2 a^2} + V(\phi)  \vv \,\,\, p_{\phi} = \frac{\phi'^2}{2 a^2} - V(\phi)  \vv \,\,\, w_{\phi} = \frac{p_{\phi}}{\rho_{\phi}} \pp \ee
Finally, we can express the conservation equations for dark energy and growing
neutrinos as follows \cite{wetterich_1995, amendola_2000}: 

\bea \label{cons_phi} \rho_{\phi}' = -3 {\cal H} (1 + w_\phi) \rho_{\phi} +
\beta \phi' (1-3 w_{\nu}) \rho_{\nu} \vv \\
\label{cons_gr} \rho_{\nu}' = -3 {\cal H} (1 + w_{\nu}) \rho_{\nu} - \beta \phi' (1-3 w_{\nu}) \rho_{\nu}
\pp \eea The sum of the energy momentum tensors for neutrinos and dark energy is conserved,
 but not the separate parts. We neglect a possible cosmon coupling
to Cold Dark Matter (CDM), so that $\label{cons_cdm} \rho_c' = -3 {\cal H} \rho_c $.

\begin{figure}[ht]
\begin{center}
\begin{picture}(185,235)(20,0)
\includegraphics[width=85mm,angle=0.]{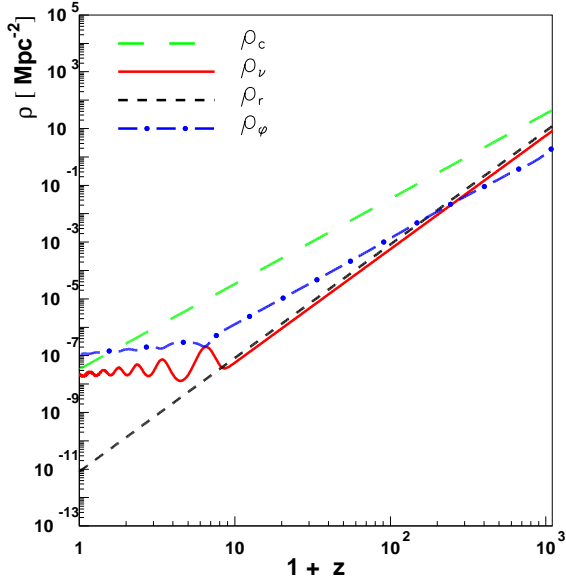}
\end{picture}
\end{center}
\caption{Energy densities of neutrinos (solid, red), cold dark matter (long-dashed, green), dark energy (dot-dashed, blue) and photons (short dashed, black) are plotted vs redshift. For all plots we take a constant $\beta = -52$, with $\alpha = 10$ and large average neutrino mass $m_\nu = 2.11 \, {\rm eV}$.}
\label{fig_1}
\vspace{0.5cm}
\end{figure}

Given the potential (\ref{pot_def}), the evolution equations for the different species can be numerically integrated, providing the background
evolution shown in fig.\ref{fig_1}. (Here we choose $\beta = -52$, $\alpha = 10$ as in the original proposal \cite{amendola_etal_2007}). The initial
pattern is a typical early dark energy model, since neutrinos are still relativistic and almost
massless, with $p_\nu = \rho_\nu/3$ so that the coupling term in eq.(\ref{kg}) and (\ref{cons_phi}, \ref{cons_gr}) vanishes. Dark
energy is still subdominant and falls into the attractor provided by the
exponential potential (see \cite{wetterich_1995, amendola_2000, copeland_etal_1998} for
details), in which it tracks the dominant background component with an early dark energy fraction $\Omega_h = n/\alpha^2$ and $n = 3(4)$ for the matter (radiation) dominated era. Radiation dominates until matter radiation equality, then CDM takes over. As the mass of the neutrinos increases with time, the coupling term
$\sim \beta \rho_\nu$ in the evolution equation for the cosmon (\ref{kg}) (or equivalently in (\ref{cons_phi}, \ref{cons_gr}))
starts to play a significant role, kicking
$\phi$ out of the attractor as soon as neutrinos become nonrelativistic. In fig.{\ref{fig_1}} this is visible in the modified behavior of $\rho_\nu$ and $\rho_\phi$ for $z < 10$. Subsequently, small decaying oscillations characterize the $\phi - \nu$ coupled fluid
and the two components reach almost constant values. The va\-lues of the energy densities
today are in agreement with observations, once the precise crossing time for the end of the scaling 
solution has been fixed by an appropriate choice of the coupling $\beta$. At
present the neutrinos are still subdominant with respect to CDM, though in the future
they will take the lead.

For completeness, note that the unperturbed neutrino pressure reads
\be
p_\nu = \frac{1}{3} a^{-4} \int{q^2  dq d\Omega \frac{q^2}{\epsilon(\phi)} f_0(q)} \vv
\ee
\\
where ${\bf q} = a {\bf p} = q {\bf \hat{n}}$ is the comoving 3-momentum, $\epsilon = \epsilon(\phi) = \sqrt{q^2 +m_\nu(\phi)^2 a^2} $, $f$ is the phase space distribution and $f_0$ its zeroth-order term (Fermi-Dirac distribution).
The neutrino energy density can either be given by solving the conservation equation (\ref{cons_gr}) or equivalently via the integral:
\be \rho_\nu = a^{-4} \int{q^2 dq d\Omega {\epsilon(\phi)} f_0(q)}  \pp \ee

\section{Nonlinear evolution equations} \label{nl}

Once neutrinos become nonrelativistic ($z \sim 5-10$), they start feeling an attractive force stronger than gravity and mediated by the cosmon field. Neutrino perturbations rapidly grow and become nonlinear at a redshift $z \sim 1-2$ \cite{mota_etal_2008}, when they might form stable lumps, whose solutions have been described in \cite{brouzakis_etal_2007}. Our intent in this section is to investigate, via a nonlinear analysis in physical space, the formation and evolution of neutrino lumps and their properties in redshift, in order to estimate the final gravitational potential characterizing the lumps as a function of their final scale.

With this aim in mind, we solve the nonlinear Navier-Stokes equations in an expanding Universe and position space with comoving coordinates $\bf{x}$:

\bea \label{eq:com_ns1} \delta_{\nu}' &=& -\scalprod{{\mathbf v}_{\nu}}{\mathbf \nabla}\delta_{\nu} - (1 + \delta_{\nu})\scalprod{\mathbf \nabla}{{\mathbf v}_{\nu}} \vv \\
\label{eq:com_ns2} {\mathbf v}_{\nu}' &=& -\left({\cal H} - \beta\phi'\right)\,\mathbf{v_{\nu}} - \left(\scalprod{\mathbf{v}_{\nu}}{\mathbf \nabla}\right)\mathbf{v}_{\nu} \nn \\
                                      &&\quad + {\bf \nabla} (\Phi_{\nu} + \beta\,\delta\phi) \vv \\
\label{eq:com_poisson} \Delta\delta\phi &=& -\beta\,a^2\,\delta_{\nu}\bar\rho_{\nu} \vv \\
\label{eq:com_grav_pot} \Delta\Phi_{\nu} &=& -\frac{a^2}{2}\,\delta_{\nu}\bar\rho_{\nu} \pp \eea
Here $\bar\rho_\nu$ is the background neutrino energy density and $\delta_\nu \equiv \delta \rho_\nu /\bar\rho_\nu$ is the relative neutrino density perturbation ($\sim 1$ when reaching nonlinearity). The vector $\bf{v}_\nu$ is the velocity for neutrinos. More precisely, it describes the peculiar comoving velocities - it vanishes for neutrinos with constant comoving coordinates. The evolution of the velocities is driven by the gradients of the gravitational potential and cosmon field, with the usual Hubble damping and quadratic term arising from particle number conservation. The velocity dependent term $\beta\phi'\mathbf{v}_{\nu}$ in eq.(\ref{eq:com_ns2}) is not present in the standard Navier Stokes equations. It accounts for momentum conservation, reflecting the fact that the neutrino mass changes in time as $m'/m = -\beta \phi'$. This friction term can be rigorously derived within the fully relativistic equations, as outlined in the last section of this paper.

Equation (\ref{eq:com_grav_pot}) is the Poisson equation for the gravitational potential that we have indexed as $\Phi_{\nu}$ to clarify that it only comprises the neutrino contribution. 
(The sign convention for the gravitational potential in eqs.(\ref{eq:com_ns2}) (\ref{eq:com_grav_pot}) matches the linear equations in \cite{mota_etal_2008} with $\Delta$ corresponding to $-k^2$ in momentum space). Eq.(\ref{eq:com_poisson}) is the perturbed Klein Gordon equation in the limit in which time derivatives are negligible with respect to the spatial ones, as it holds in the Newtonian approximation.
Effectively, eq.(\ref{eq:com_ns1}) relates the time evolution of the neutrino overdensity $\delta_\nu$ to the divergence of its corresponding momentum density. The time dependence of the momentum density itself is directly connected, via eq. (\ref{eq:com_ns2}), to the given forces: in addition to the gravitational force $\vec{F}_{g} = {\vec \nabla} \Phi_{\nu}$, the cosmon mediated fifth force $ \label{eq:fifth_force_def} \vec{F} = \beta \vec{\nabla}\delta\phi $ is present, in the form derived already at a linear perturbation level (see sec. \ref{sec:rel_nonrel}). Combining eqs.(\ref{eq:com_poisson}) and (\ref{eq:com_grav_pot}) immediately shows that $\delta \phi \sim 2 \beta \Phi$, clarifying that the scalar field mediates a force of order $|\vec{F}| = |\beta \vec{\nabla}\delta\phi| \sim 2\beta^2 |\vec{F}_g|$. For a choice of $\beta \sim -50$ this is about $5000$ times stronger than gravity.


In section \ref{sec:rel_nonrel} we will show that eqs.(\ref{eq:com_ns1}) and (\ref{eq:com_ns2}) can be obtained by considering the appropriate limits of the fully relativistic equations derived from the Bianchi identity in presence of an external source \cite{kodama_sasaki_1984}
\be
\label{eq:ps_cons}\nabla_{\gamma}T_{\mu}^{\gamma} = Q_{\mu} = -\frac{\beta}{M}  T_{\gamma}^{\gamma}  \partial_{\mu}\phi \vv
\ee
where $T^{\gamma}_{\mu}$ is the stress energy tensor of the neutrino fluid. 

For convenience we display eqs.(\ref{eq:com_ns1} - \ref{eq:com_grav_pot}) also in terms of cosmic time $t$ and physical (not comoving) coordinates:
\bea \label{eq:com_ns1_t} \frac{\partial\rho_{\nu}}{\partial t} &=& -\nabla(\rho_\nu\mathbf{v_{tot}}) - \beta\dot\phi\,\rho_{\nu} - \beta\,\nabla\phi\,\rho_{\nu}\,\mathbf{v_{tot}} ,  \\
\label{eq:com_ns2_t} \frac{\partial{\mathbf v}_{tot}}{\partial t} &=& \left(\nabla\mathbf{v_{tot}} + \beta\,\dot\phi\right)\mathbf{v_{tot}} + \nabla\left(\Phi_{\nu} + \beta \phi\right)  \\ &-& \left(\scalprod{\mathbf{v}_{tot}}{\mathbf \nabla}\right)\mathbf{v}_{tot} \vv \nn \eea
Here $\rho_\nu$ and $\phi$ are the local neutrino density and cosmon fluid including the fluctuations and $\bf{v_{tot}}$ is the total neutrino velocity, composed of Hubble flow and peculiar velocity as  \be {\bf{v}_{tot}} = \frac{d {\bf{r}} }{dt} = H {\bf r} + {\bf {v}_\nu} \pp \ee
Equations \eqref{eq:com_ns1_t} and \eqref{eq:com_ns2_t} may be combined to yield the conservation equation of the momentum density ${\mathfrak p} \equiv \rho_\nu \mathbf {v}_{tot}$, namely
\be \label{eq:com_moment_t} \dot{\mathfrak p}_i = {\bf f}_{\text{disp}(i)} + {\bf f}_{\text{attr}(i)} \vv \ee
where we have defined
\bea{\bf f}_{\text{disp}(i)} &\equiv& - \partial_j\left[\rho_\nu {\mathbf v}_{tot(i)}{\mathbf v}_{tot(j)}\right] \vv \label{fdisp} \\
{\bf f}_{\text{attr}(i)} &\equiv& \rho_\nu \partial_i(\Phi_{\nu} + \beta\, \phi) \vv \nn
\eea
The attractive force ${\bf f}_{\text{attr}}$ may eventually be balanced by the countering force associated to the velocity dispersion, ${\bf f}_{\text{disp}} $. We have omitted in eq.(\ref{fdisp}) and eq.(\ref{eq:com_ns1_t}) an additional pressure force \bea {\bf f}_{\text{press}(i)} &\equiv& -3(H-\beta \dot{\phi}) p_{\nu} {\bf v}_{tot(i)} \vv \\ p_\nu &=& w_\nu (\rho_\nu + \delta \rho_\nu) \nn \vv \eea
which is important only when neutrinos are still relativistic. It prevents the growth of neutrino overdensities for $z \gsim 10$.

\section{Evolution of neutrino lumps} \label{sec:results}

\subsection{Method} We integrate the set of nonlinear equations (\ref{eq:com_ns1}-\ref{eq:com_grav_pot}) on a 128x128x128 point 3-dimensional spatial grid of fixed size $L^3$, by means of a method of lines with an adaptive time stepper. The length $L$ is chosen to match the given initial profile, such that a good resolution around its maximum can be obtained. We impose cubic periodic boundary conditions, $\rho(\mathbf{x}) = \rho(\mathbf{x} + \mathbf{L})$ with $L_i = 0,L$. For each time step, the Poisson equations (\ref{eq:com_poisson}) (\ref{eq:com_grav_pot}) are solved employing a Fast Fourier Transform routine.
 
We start by considering the formation of a single lump within our box. As an initial density profile we consider a gaussian: 
\be
\label{eq:dens_ini} \delta_{\nu}(\mathbf{x}) = h_{in}e^{-\mathbf{x}^2/r_{in}^2} \pp
\ee
The initial density amplitude $h_{in}$ is chosen to match the corresponding linear overdensities at the matching redshift $z_{m}$, an issue which will be discussed in more detail in the next section. 
A convenient width of the box corresponds to about $L \sim 6r_{in}$. 
We start the numerical integration early enough ($z \sim 9$) in order to allow us to start with rather arbitrary small neutrino velocities $\mathbf{v}_{in}$ (see below). Independently of their precise values there is enough time for their radial component to adapt such that linear perturbations are matched by the time we reach the range of validity of the nonlinear equations specified above. We also investigate alternative initial conditions for the velocities. The dependence of our results on initial conditions is discussed in the next paragraph, while illustrating our results.

\subsection{Turn over and free fall}
One of our aims is to understand and illustrate the evolution of neutrino lumps in time: we expect that after a first phase in which the lump expands with the background, peculiar velocities will take over. The lump will then decouple from the background expansion and contract with increasing inward velocities. A useful quantity to illustrate the rapid increase in velocity is the kinetic energy of the lump, defined in physical coordinates as
\be \label{eq:T} \text{E}_{\text{kin}} = \frac{1}{2}\,\int_V \delta\rho_{\nu}\,\mathbf{v}_{\text{tot}}^2\,d^3x \ee
with total velocity $\mathbf{v}_{\text{tot}} = \mathbf{v}_\nu + {\cal H}\mathbf{x} = \mathbf{v}_\nu + H\,\mathbf{r}$, being the sum of peculiar and expansion contributions. Comparing the kinetic contribution to the potential energy can give us an indication of when and how fast the lump approaches virialization. The potential energy is defined as 
\be \label{eq:V} \text{E}_{\text{pot}} = - \frac{1}{2}\,\beta \int_V \delta\rho_{\nu}\,\delta\phi\,d^3x \ee
where $V$ denotes the physical volume of the overdensity. We have omitted here the small gravitational potential energy. Both $E_{kin}$ and $E_{pot}$ are evaluated in practice in cartesian coordinates rather than in spherical ones, as late phases of the collapse might involve important anisotropies. 

\begin{figure}[ht]
\begin{center}
\includegraphics[width=85mm,angle=-0.]{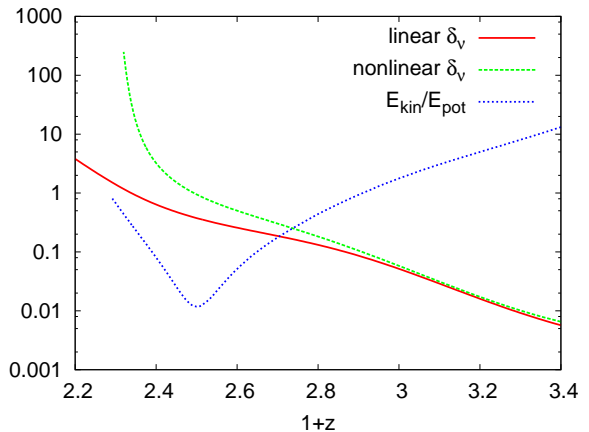}
\\
\end{center}
\caption{Linear (solid, red) and nonlinear (long-dashed, green) neutrino overdensity $\delta_\nu$ vs redshift, until virialization occurs. The nonlinear overdensity is evaluated in the center of the lump. The ratio of kinetic over potential energy associated to the lump is also shown (short-dashed, blue). The comoving (physical) initial lump radius $r_{in} = 45$ Mpc ($4.46$ Mpc) fixes the box size $L = 270$ Mpc for the simulation and corresponds roughly to a scale $k/h = 0.1 \text{Mpc}^{-1}$ for $h = 0.72$. It also determines a final physical radius of the lump $R_f = 3.21 r_{in} = 14.3$ Mpc.}
\label{fig_virialplot}
\vspace{0.5cm}
\end{figure} 

\begin{figure}[ht]
\begin{center}
\includegraphics[width=85mm,angle=-0.]{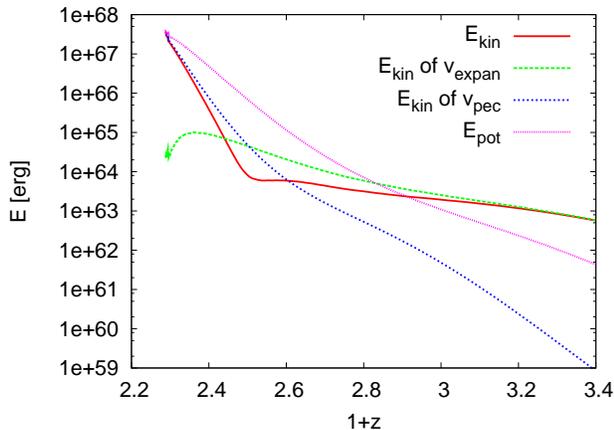}
\\
\end{center}
\caption{Evolution of potential (pink, dotted) and kinetic energy (red, solid) versus redshift. The two main contributions to the kinetic energy are also shown: the energy associated with the Hubble expansion is depicted in green (long-dashed) and the one due to peculiar motion in blue (short-dashed). The size of the lump is the same as in fig.\ref{fig_virialplot}. An energy of $10^{66}$ erg corresponds to roughly $6\cdot10^{77}$ eV.}
\label{fig:e_kin_pot}
\vspace{0.5cm}
 \end{figure} 

The nonlinear evolution of a perturbation as resulting from the integration described above is shown in Fig.\ref{fig_virialplot}. It is compared to the corresponding linear evolution. We have chosen $r_{in} = 45$ Mpc ($4.46$ Mpc) in comoving (physical) units and find a characteristic final size of the lump $R_f = 3.21 r_{in} = 14.3$ Mpc in physical units. The precise definition of $R_f$ will be given below. The ratio of kinetic to potential energy is also shown in the plot. The evolution of the potential energy and of the kinetic energy, together with the split of the kinetic energy into expansion and peculiar components, is further shown in Fig.\ref{fig:e_kin_pot} for the same final scale. For the quantitative evaluation of the energies we choose a volume that extends to a radius where the density contrast reaches $(1/50)$ times the central density contrast.

By looking at Fig.\ref{fig_virialplot} and Fig.\ref{fig:e_kin_pot}, we can identify three redshift ranges:
\begin{description}
  \item{[i]} $1+z > 3$: In this range, density perturbations are still linear; total velocities and therefore kinetic energy are dominated by the expansion term $\delta\rho_{\nu}{\cal H}^2\mathbf{x}^2 (= \delta\rho_{\nu}{H}^2\mathbf{r}^2$) present in eq.(\ref{eq:T}) while the peculiar velocities $\mathbf{v}$ are negligible. At this stage, both potential and kinetic energy are increasing due to the accretion of more neutrinos in the lump and to the increase in the neutrino mass. The potential energy $E_{pot}$ increases somewhat faster (both $\delta \rho_\nu$ and $\delta \phi$ are proportional to $m_\nu$), which results in a decreasing ratio $E_{kin}/E_{pot}$. 
\item{[ii]} $2.5 < 1+z < 3$: At $z\sim 2$ the attractive fifth force starts to become the dominant contribution in eq.(\ref{eq:com_moment_t}). In this regime we can trust the nonlinear evolution for the chosen scale. The nonlinear density perturbation detaches from the linear one. Also the radial peculiar velocities start to become nonnegligible, adding an inward contribution to the outward expansion term. As a consequence, for $1+z<3$ the increase of the kinetic energy is slowed down. The latter effect is more pronounced for inner shells of the profile, for which the radial velocities change sign, while outer shells still tend to expand with the background Universe. During this period $E_{pot}$ keeps increasing, with a slightly steeper slope than in the previous range due to the effects of nonlinearity on the neutrino density perturbation.
\item{[iii]} $2.3 < 1+z < 2.5$: The nonlinear effects on the density perturbation become very pronounced. Peculiar velocities dominate over the expansion term and induce a rapid increase of the kinetic energy, due to the inward velocities. As a consequence, the characteristic scale of the lump shrinks until virialization is reached. Equation (\ref{eq:com_ns2}) is now completely dominated by the fifth force. The lump has effectively decoupled from the background and evolves according to the laws of free fall. At the end of this period the tangential peculiar velocities as well as irregularities in the spherical symmetry of the lump grow rapidly. The borders of numerical precision are reached. The latter effects lead to a highly anisotropic behavior, gradients become very large and velocities change directions almost randomly. The timestepper eventually stops the integration. Note that our code stops as the ratio $|E_{kin}/E_{pot}|$ approaches the value of $1/2$, corresponding to the virialization condition. The latter has not been introduced by hand in the equations but is automatically reached due to increasing tangential peculiar motions and deviations from spherical symmetry.
 \end{description}

\begin{figure}[ht]
\begin{center}
\includegraphics[width=85mm,angle=-0.]{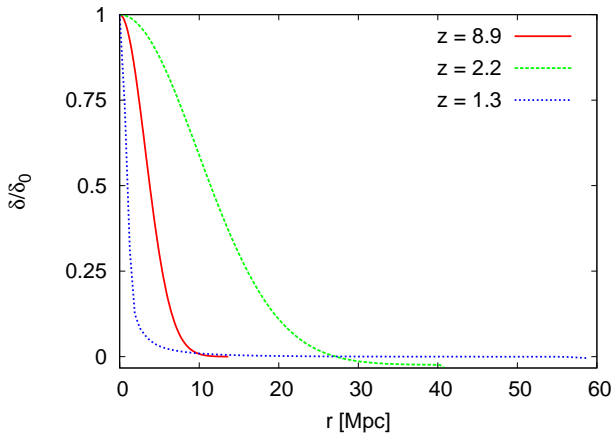}
\end{center}
\caption{Neutrino overdensity profiles at redshifts z = 8.9 (red, solid), z = 2.2 (green, long-dashed) and z = 1.3 (blue, short-dashed) in terms of physical distance $r$ from center, averaged over all angles. The central overdensities are $\delta_0 = \delta(0,z) = 1.1\cdot10^{-5},\,1.6\cdot10^{-2},\,250$, respectively. To illustrate the change in shape we have normalized the profiles by dividing out $\delta_0$. The radius is given in physical units. An initial broadening due to the expansion is followed by a strong concentration process. An apparent discrepancy between the width of the profiles and the radii shown in fig.\ref{fig:R} is caused by the different definition (\ref{eq:char_scale}).}
\label{fig:dens_profile}
\vspace{0.5cm}
\end{figure}

\begin{figure}[ht] 
\begin{center}
\includegraphics[width=85mm,angle=-0.]{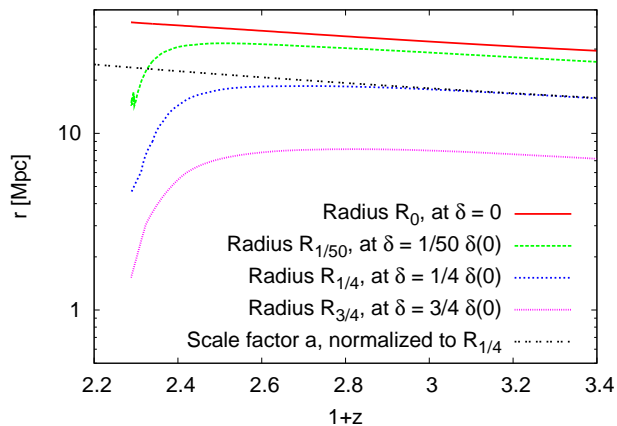}
\\
\end{center}
\caption{Radii of the neutrino overdensity profile at different amplitudes, for a final maximal radius of $R_f = 14.3$ Mpc. We plot the time evolution of the physical radii where the neutrino overdensity reaches a certain fraction of the core density.}
\label{fig:R}
\vspace{0.5cm}
\end{figure} 
\begin{figure}[ht] 
\begin{center}
\includegraphics[width=85mm,angle=-0.]{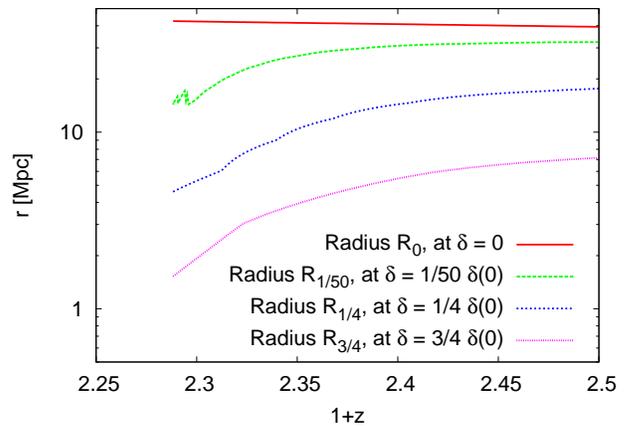}
\\
\end{center}
  \caption{Late evolution of different radii for a final maximal radius of $R_f = 14.3$ Mpc. This plot zooms fig. \ref{fig:R} to a smaller range of $z$, such that the more irregular behavior close to virialization becomes visible.}
\label{fig:R_late}
\vspace{0.5cm}
\end{figure} 

As the neutrino lump undergoes the different evolutionary stages, the density profile adapts according to our equations once the nonlinear regime is reached. While density perturbations are still linear, the profile expands with the background, leaving its shape almost inaltered. The only major change is the buildup of a surrounding underdensity as neutrinos flow into the lump. Note that for a single lump the underdensities are always much smaller than the background density, even when the overdensities become large. When collapse is approached, the profile changes its shape. As outer shells turn around at smaller redshifts, the slope of the profile changes. Its initial gaussian shape changes into a profile which resembles common density profiles for dark matter halos, such as the Navarro-Frenk-White profile. We plot the density profile for different redshifts in fig.{\ref{fig:dens_profile}}.

The size of the neutrino lump is the most characteristic quantity which distinguishes the evolution of different types of inhomogeneities. If we start with vanishing initial peculiar velocities and adjust $h_{in}$ ($r_{in}$) according to the linear evolution of small perturbations, our initial conditions depend on only one relevant parameter $r_{in}$. We are interested to determine the final size $R_f$ of the neutrino lump in terms of $r_{in}$. An understanding of the time evolution of the different length scales in the lump is also important for the precise definition of the volume integral in eqs.(\ref{eq:T}) (\ref{eq:V}).

It is therefore interesting to consider different radii of the profile, characterizing shells at different distance from the center of the lump. In order to define radii corresponding to inner or outer shells, we consider the physical radius $R(\lambda, z)$ for which the amplitude reaches a given fraction $\lambda$ of the central amplitude: $\delta(z, R(\lambda,z)) \approx \lambda \delta(z, 0)$. The radius $R(\lambda, z)$ depends on the redshift $z$. Higher values of $\lambda$ correspond to amplitudes which are closer to the central value and therefore to inner shells inside the lump. Since the lump has no spherical symmetry we have to make the definition of $R(\lambda, z)$ more precise by taking the largest value of $\vec{r}$ for which the density $\lambda \delta(0)$ is reached, or formally define the radius $R(\lambda, z)$ as the superior limit of the $\{\vec{r}\}$ ensemble. 
In formulae we have:
\be \label{eq:char_scale} R(\lambda,z) := sup\{\,|\vec{r}| \,\,\,\, | \,\,\,\, \delta_{\nu}(z,\vec{r}) \geq \lambda\,\delta_{\nu}(z,0)\} \pp \ee
Here the factor $\lambda$ is a constant $0 \le \lambda \le 1$ and all spatial coordinates $\vec{r}$ are again physical.

We show the time evolution of $R(\lambda, z)$ in units of $r_{in}$ in Figs.\ref{fig:R} and \ref{fig:R_late}. We have chosen four different values for $\lambda$, i.e. $\lambda \in \{0,1/50,1/4,3/4\}$ and display $R_0(z) = R(\lambda = 0,z)$, $R_{1/50}(z) = R(\frac{1}{50},z)$, $R_{1/4}(z) = R(\frac{1}{4}, z)$ and $R_{3/4}(z) = R(\frac{3}{4},z)$. As expected, for larger values of $\lambda$ the corresponding shell turns around earlier. For comparison, we have also plotted the slope corresponding to the scale factor $a(z)$. As expected it corresponds to the slope of $R_\lambda$ within the redshift regime in which the expansion is still dominant. As a measure of the size of the lump we employ $R_l(z) = R_{1/50}(z).$ We assume that the change of $R$ after virialization can be neglected, such that the final size $R_f$ is given by $R_f = R(z_{vir}) = R(1/50, z_{vir}).$

\subsection{Onset of virialization}

\begin{figure}[ht] 
\begin{center}
\includegraphics[width=85mm,angle=-0.]{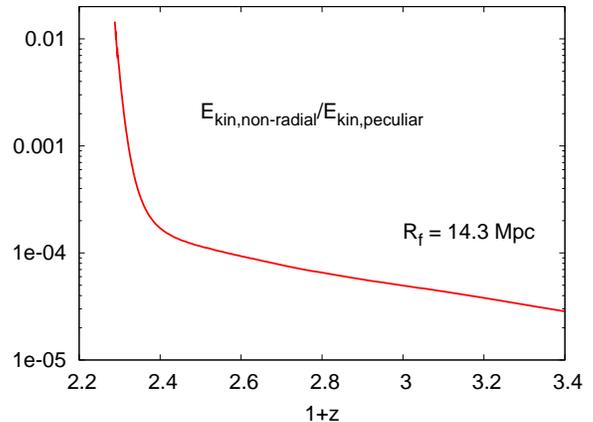}
\\
\end{center}
\caption{Ratio of the nonradial part of the peculiar kinetic energy and peculiar kinetic energy as a function of redshift.}
\label{fig:nonradial}
\vspace{0.5cm}
\end{figure}

The late evolution of the different radii displayed in Fig.\ref{fig:R_late} shows a rather irregular behavior. This indicates the onset of virialization, as also suggested by the ratio $E_{kin}/E_{pot}$ in Fig.(\ref{f2}). The reason is that transversal velocities become important in this redshift range. In order to illustrate this we
analyze the evolution of nonradial velocities within the lump. In fig. \ref{fig:nonradial}, we plot the ratio $E_{\text{kin, nonradial}}/E_{\text{kin,peculiar}}$, with

\bea \label{eq:e_kin_nonrad} E_{\text{kin, nonradial}} &=& \frac{1}{2}\,\int_V \delta\rho_{\nu}\,\delta\mathbf{v}_{\nu}^2\,d^3x \vv \\
\label{eq:e_kin_pec} E_{\text{kin,peculiar}} &=& \frac{1}{2}\,\int_V \delta\rho_{\nu} \mathbf{v}_{\nu}^2\,d^3x \vv \eea
where $\delta \mathbf{v} \equiv \mathbf{v} - (R'/R) \mathbf{r}$, $\mathbf{v}$ is the peculiar velocity, $R$ the radius of the lump and $r$ is the radial coordinate.
While for higher redshifts the nonradial contribution to the peculiar kinetic energy is entirely negligible, it increases rapidly once the evolution of the lump has predominantly turned around.

In order to understand this, it is useful to consider a shell with a given radius $R_s(\tau)$ and analyze equation (\ref{eq:com_ns1}) to see how deviations from its mean velocity evolve \cite{engineer_et_al_2000}. For this purpose, we transform into a coordinate system in which the mean of the shell is at rest, $\mathbf{\tilde{x}} = {R_s}^{-1}(\tau)\,\mathbf{r} = (a/R)\,\mathbf{x}$, where $\mathbf{r}$ and $\mathbf{x}$ are the physical and comoving spatial coordinates, respectively. We then decompose the peculiar velocities into a radial part ${\mathbf v}_{||} = \mathbf{x}'$ and a nonradial part $\delta\mathbf{v_s}$ and insert it into equation (\ref{eq:com_ns2}). Appropriately transforming spatial derivatives leads to the expression. 
\bea \label{eq:v_aniso} \delta\mathbf{v}_{s}' &=& -\left(\frac{R'_s}{R_s} - \beta\phi'\right)\,\delta\mathbf{v}_{s} \\ &&- \frac{a}{R_s}\,\left[(\scalprod{\delta{v}_{s}}{\nabla})\,\delta\mathbf{v}_{s} - \,\nabla_{\perp}(\Phi_{\nu} + \beta\,\delta\phi) \right] \pp \nonumber \eea
The gradient $\nabla_\perp$ in front of the potentials shows that only nonradial components influence the evolution. This may also be understood by noting that the underlying situation is formally equivalent to the evolution of the velocity perturbation within a Universe expanding with the rate $R'/R$. The structure should thus simply resemble equation (\ref{eq:com_ns2}) with an altered Hubble function. The additional factors of $a/R$ are due to the choice of time coordinate.

As long as the shell is expanding, the friction term $R'/R$ behaves as a damping force. Once the shell turns around, however, $R'$ becomes negative, and the term effectively enhances the build-up of nonradial velocities. Due to the large modulus of the collapse rate, this happens quite rapidly.
Of course, an exactly radial flow will remain radial by virtue of rotation symmetry (up to numerical errors). We have therefore started initially with small anisotropies in the velocity (or density) distribution. For this purpose, we have imposed an additional initial velocity field ${\bf v}_{rand}$, whose directions were randomly distributed. Its amplitude at each space point $|{\bf v}_{rand}|$ was chosen to be of an order of a few percent of the mean initial expansion velocities. We find that the late stages of the flow are independent of the precise choice of initial conditions for the nonradial velocities.

\section{Gravitational potential of neutrino lumps}
\subsection{Single lump potential}
\begin{figure}[ht] 
\begin{center}
\includegraphics[width=85mm,angle=-0.]{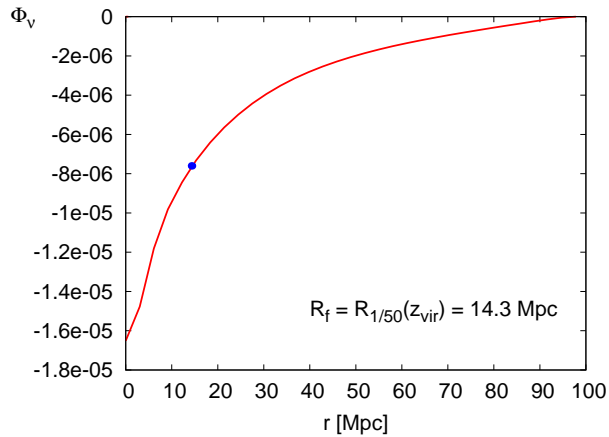}
\\
\end{center}
\caption{Dependence of gravitational potential on distance from the center at redshift of virialization $z = 1.3$. The dot indicates $R_f$. }
\label{fig:phin_r}
\vspace{0.5cm}
\end{figure}

\begin{figure}[ht]
\begin{center}
\includegraphics[width=85mm,angle=-0.]{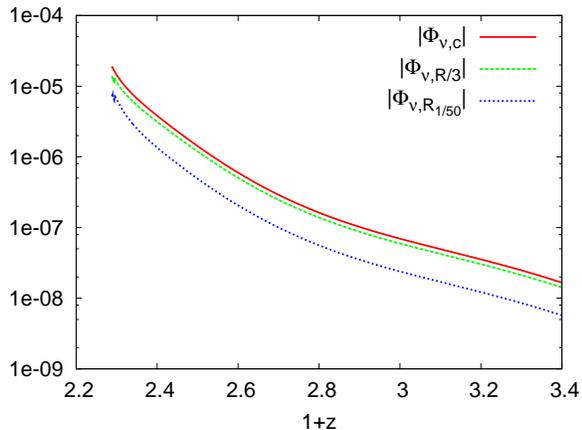}
\\
\end{center}
\caption{Neutrino gravitational potential $\Phi_{\nu}$ as a function of redshift for a fixed final scale $R_f = 14.3$ Mpc. We evaluate $\Phi_{\nu}$ for different distances from the center of the lump, namely at $R_l = R_{1/50}$, $R_l/3$ and in the center of the lump $(\Phi_{\nu,c})$.}
\label{fig:grav_pot_vs_z}
\vspace{0.5cm}
\end{figure}

The key quantity of interest is the characteristic gravitational potential of a neutrino lump. This will influence the peculiar velocities of galaxies or gas, or the CMB-anisotropies via the ISW effect. Of course, the gravitational potential depends on the distance from the center of the lump as shown in fig.{\ref{fig:phin_r}}. We have indicated the scale $R_f = R_{1/50}(z_{vir})$ by a dot. Fig.\ref{fig:phin_r} demonstrates that the radius $R_{1/50}$ is a typical scale which characterizes the gravitational potential of the lump. We may define $\Phi_{\nu,0}$ as the gravitational potential in the center of the lump, $\Phi_{\nu,R_l}$ as its value at the radius $R_l$ defined in eq.(\ref{eq:char_scale}), and $\Phi_{\nu,R_l/3}$ a corresponding value at distance $R_l/3$ from the center. We plot in Fig.\ref{fig:grav_pot_vs_z} the time evolution of the gravitational potential, for a lump with final radius $R_f = 14.3$ Mpc. Note again that $\Phi_{\nu}$ is obtained as the solution to the Poisson equation (\ref{eq:com_grav_pot}).

\begin{figure}[ht]
\begin{center}
\includegraphics[width=85mm,angle=-0.]{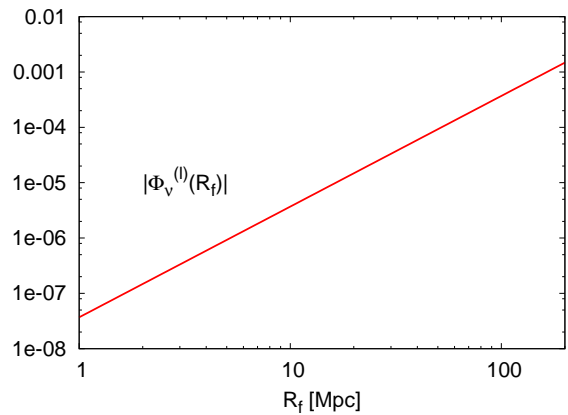}
\end{center}
\caption{Gravitational potential as a function of the final size $R$ of the neutrino lump (at redshift $z=0$). We define $R$ as the radius at virialization $R_f$, assuming that it remains approximately constant from virialization up to now.}
\label{fig:grav_pot_of_scale}
\vspace{0.5cm}
\end{figure}

In fig.\ref{fig:grav_pot_of_scale} we consider families of lumps with different initial conditions and show the gravitational potential of the lump versus its final size $R_f$. We plot the nonlinear value of the gravitational potential $\Phi_{\nu}$ as obtained from the numerical solution of eqs.(\ref{eq:com_ns1}-\ref{eq:com_grav_pot}) for the definition of $R_f = R(1/50,z_{vir})$ according to eq. (\ref{eq:char_scale}). This value has to be handled with care. We have investigated in this paper only the situation where a single lump forms, essentially independently of possible surrounding lumps. In our numerical work, this follows from the choice of initial conditions. In the true Universe, the situation may be much more complicated. Many lumps of different sizes are expected to form, competing for the available neutrinos. Even though these ``initial lumps'' may eventually merge to bigger structures, such a merging process can be very different from the essentially spherical infall investigated in this paper. One typically expects a smaller gravitational potential for a big structure arising from merging as opposed to spherical infall, since the ``substructures'' carry tangential velocities. In this respect our result for $\Phi_{\nu}$ should be considered as an upper bound, in particular for large values of $R$. It should become a good approximation for small enough $R$ where the possible merging processes do not play a dominant role. Without extended numerical simulations it is difficult to assess the value of $R$ beyond which the true characteristic gravitational potential remains substantially below the values in fig.\ref{fig:grav_pot_of_scale}. With these words of caution the final value for a virialized structure of typical size $\sim 10 (100)$ Mpc is $\Phi_\nu \sim 10^{-6} (10^{-4})$.

\subsection{Neutrino induced cosmological gravitational potential}
For an estimate of the cosmological impact of the gravitational potential of the neutrino lumps one needs a relation between the characteristic potential for a single lump and the average cosmological value of the gravitational potential at a given scale, expressed as the Fourier component $\Phi^{(c)}_{\nu,k}$. One expects a typical suppression factor $\gamma_c$ for the cosmological gravitational potential as compared to the potential of a single lump with size $R_f = \pi/k$, 
\be \label{eq:gamma_c} \gamma_c(k) \equiv \frac{|\Phi^{(c)}_{\nu,k}|}{|\Phi^{(l)}_{\nu,R}|}\,,\,\,\,k = \frac{\pi}{R} \pp \ee
For clarity we here indicate the cosmological potential and the single lump potential by superscripts $(c)$ and $(l)$. If all lumps would have the same size, the suppression should be roughly given by the fraction of the volume occupied by neutrino lumps.
For a rough numerical estimate of $\gamma_c$ we may then use the relations 
\be \bar\rho^{(l)}_{\nu} V_\text{lumps} = N_{\nu}^{\text{(lumps)}}\,,\,\,\,\rho_{\nu}^{(c)} V_\text{hor} = N_{\nu} \vv \ee
with $V_\text{lump}$ the total volume occupied by neutrino lumps and $V_\text{hor}$ the volume within the cosmological horizon, $V_\text{hor} \approx \frac {4\pi}{3} \left(3000\text{ Mpc}\right)^3$, and define by $F^{(l)}_{\nu} \equiv N_{\nu}^{\text{(lumps)}}/N_{\nu}$ the total fraction of neutrinos which are bound in lumps of size $R = \pi/k$. This estimates $\gamma_c$ in terms of the ratio of the average energy density $\bar\rho_{\nu}^{(l)}$ of the neutrinos in the lumps and the cosmological background density $\rho_{\nu}^{(c)}$
\be \label{eq:gamma_c_est} \gamma_c \approx \frac{V_\text{lumps}}{V_\text{hor}} = F^{(l)}_{\nu}\,\frac{\rho_{\nu}^{(c)}}{\bar\rho_{\nu}^{(l)}} \pp \ee
For a typical neutrino density averaged over the volume of the lump within radius $R = R_{1/50}$ we find $\bar\rho_{\nu}^{(l)}/\rho_{\nu}^{(c)} \approx 100$. In our simulation we find typical values for $F^{(l)}_{\nu}$ for single lumps in a box to be $F^{(l)}_{\nu} \approx 0.1$ at $z_{vir}$; but this depends of course on the size of the box. Even if all neutrinos are finally bound in some lump one typically has $F^{(l)}_{\nu}(k) < 1$ since not all are in lumps of size $\pi/k$. We will take $F^{(l)}_{\nu} = 1/4$.

\begin{figure}[ht]
\begin{center}
\includegraphics[width=85mm,angle=-0.]{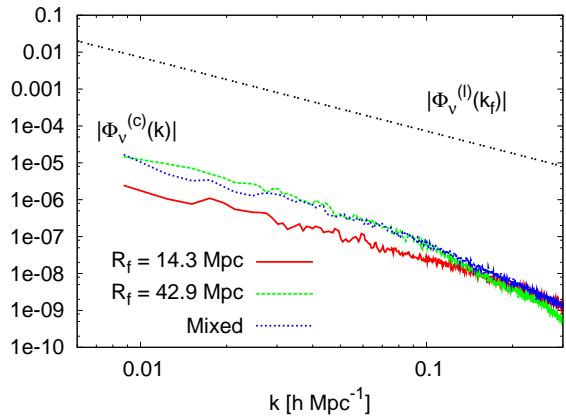}
\end{center}
\caption{Gravitational potential as obtained by Fourier transforming distributions of lumps of given size. For comparison we have included the single lump gravitational potential at distance $R = R_{1/50}$ from the center.}
\label{fig:grav_pot_FT}
\vspace{0.5cm}
\end{figure}

\begin{figure}[ht]
\begin{center}
\includegraphics[width=85mm,angle=-0.]{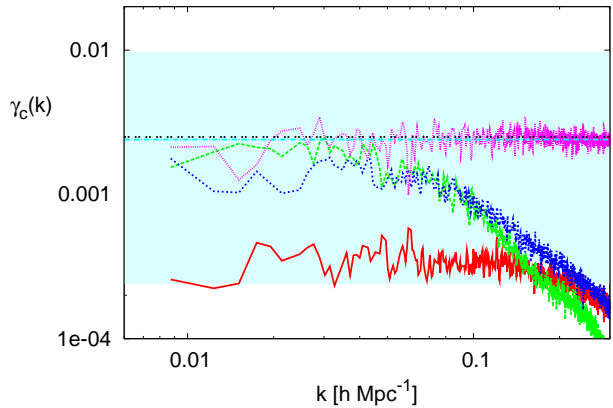}
\end{center}
\caption{Suppression factor $\gamma_c(k)$ as obtained by using the definition (\ref{eq:gamma_c}) for distributions of lumps of size $R_f = 14.3$ Mpc (solid, red), $R_f = 63.6$ Mpc (long-dashed, green) and a mixed distribution (short-dashed, blue). The dotted pink line marks the gravitational potential of the mixed distribution of point masses, while the dot-dashed light blue line corresponds to a suitable interpolation. The estimate (\ref{eq:gamma_c_est}) is also shown (double-dashed, black). It almost coincides with the interpolation of the mixed point mass distribution. The error bounds displayed (shaded, light blue) correspond to $4\gamma_c$ and $0.1\gamma_c$.}
\label{fig:gamma_c_of_k}
\vspace{0.5cm}
\end{figure}

For an alternative rough estimate of $\gamma_c$ we have randomly distributed virialized neutrino lumps with a final profile according to our numerical solution over a cosmological volume $V_c$. The number of lumps was chosen such that the total number of bound neutrinos matches $F^{(l)}_{\nu} N_{\nu}$. The corresponding gravitational potential $\Phi^{(c)}_\nu(\bf{x})$ has then been Fourier transformed in order to extract $\Phi^{(c)}_{\nu,k} = \frac{1}{V_c}\int d^3x\,\Phi^{(c)}_\nu({\bf x})\exp{(-i{\bf kx})}$. We show in fig.\ref{fig:grav_pot_FT} oversimplified distributions where all lumps correspond to a single size $R$. Nonetheless, using eq. (\ref{eq:gamma_c}), the order of magnitude of $\gamma_c$ can be extracted from fig.\ref{fig:grav_pot_FT}. We show the resulting $\gamma_c(k)$ in fig.\ref{fig:gamma_c_of_k}, together with the estimate (\ref{eq:gamma_c_est}), for $F^{(l)}_{\nu} = 1/4$. We also show a mixed distribution of lumps with final radii of $R_f = 14.3$ Mpc, $31.8$ Mpc and $63.6$ Mpc, with equal numbers of neutrinos in each sort of lumps and a total $F^{(l)}_{\nu} = 1/4$, as well as an analogous mixed distribution where the mass of the lumps is concentrated in the center (point masses).

A suitable interpolation of fig.\ref{fig:gamma_c_of_k}, as indicated by the dot-dashed light blue line, corresponds to a simple fitting formula
\bea \label{eq:grav_pot_fit} &\left|\Phi_{\nu}^{(c)}(k)\right| = 10^{-6} \left(\frac{k}{k_0}\right)^{-\kappa}& \vv \\
&k_0 = 3.9\cdot10^{-2} \text{ h/Mpc}\,,\kappa = 2& \pp \nn \eea
This fit should only be taken as a useful order of magnitude estimate. Fig.\ref{fig:gamma_c_of_k} illustrates that the final order of magnitude of the gravitational potential is quite sensitive to the choice of distribution. The expected cosmological distribution of neutrino lumps involves a distribution over a substantial range of lump sizes. Moreover, large lumps may have substructures. This makes it hard to estimate the relevant number $N_{\nu}^{\text{(lumps)}}(k)$ without explicit cosmological simulations of structure formation. On the one side, the total fraction of neutrinos found in lumps may come close to one. On the other hand, the distribution of the available neutrinos over lumps with various sizes may reduce the effective fraction $F^{(l)}_{\nu}$ relevant for a given scale. Especially for small $k$ near the horizon the clumping may not have had enough time to bind a large fraction of neutrinos. Small $F^{(l)}_{\nu}(k)$ would result in a further suppression of the effective $\gamma_c(k)$. Furthermore, note that our fit is likely to overestimate the power on small scales, i.e. for large $k$. At these scales, the exact shape of the profile is still of substantial importance, as can be seen from fig.\ref{fig:gamma_c_of_k}. However, the simple fitting formula (\ref{eq:grav_pot_fit}) with $\kappa = 2$ is a good estimate at large scales and therefore a viable mean to illustrate the cosmological implications of neutrino lumps.

\begin{figure}[ht]
\begin{center}
\includegraphics[width=85mm,angle=-0.]{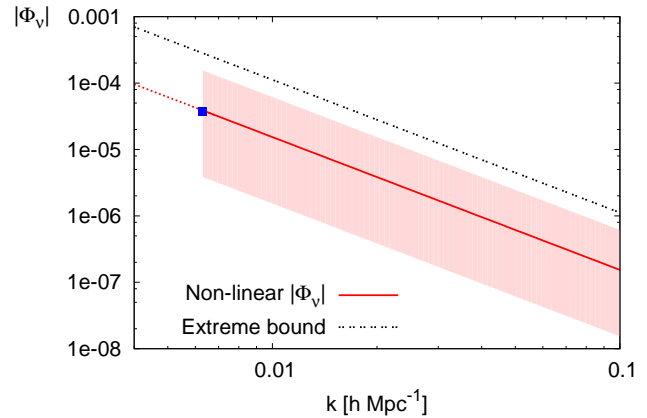}
\end{center}
\caption{Gravitational potential as a function of the scale $k$ obtained from the nonlinear (red, solid) evolution. The error bounds (shaded, red) arise from the bounds for $\gamma_c$. For scales $k$ smaller than $k_{PN}$ (solid square) the Newtonian approximation is no longer valid and the extrapolated nonlinear potential is drawn as a dotted line. The dot-dashed black line corresponds to the extreme bound (\ref{eq:Phi_up_b}).}
\label{fig:grav_pot_of_k}
\vspace{0.5cm}
\end{figure}

In fig.\ref{fig:grav_pot_of_k} we plot $|\Phi_{\nu,k}^{(c)}|$ as a function of $k$ according to eq. (\ref{eq:grav_pot_fit}). 
The substantial uncertainties in this estimate are reflected by error bounds which correspond to a multiplication of $\gamma_c(k)$ by factors of $4$, corresponding to $F^{(l)}_{\nu} = 1$, and $0.1$. The upper limit may be regarded as a rather solid bound - effects of merging dynamics and $F^{(l)}_{\nu} < 1$ reduce $\Phi_{\nu}(k)$ as compared to this bound. A more reliable estimate of $\Phi_{\nu,k}^{(c)}$ needs the actual distribution of neutrino lumps as a function of size and mass - this may perhaps require extended {\emph N}-body simulations. In fig.\ref{fig:grav_pot_of_k} we have also included the strongest possible upper bound for the cosmological gravitational potential, obtained by assuming all neutrinos in the horizon are bound within a single, point-like lump. This results in the following expression:
\bea \label{eq:Phi_up_b} \Phi^{(c)}_{\nu}(k) &=& \frac{\bar\rho_{\nu}(t_0)}{2 k^2} = m_{\nu}(t_0)\frac{n_{\nu}(t_0)}{2 M^2 k^2} \\
&=& \frac{m_{\nu}(t_0)}{2\text{ eV}}\,\frac{1.07\cdot10^{-8} (\text{h/Mpc})^{2}}{k^2} \pp \nonumber \eea
Any physically realistic scenario will lead to substantially lower gravitational potentials.

For scales $R > R_{PN} \sim 500$ Mpc, $k/h < 6.3\cdot10^{-3}$ Mpc$^{-1}$, the Newtonian approximation breaks down and our estimate of the nonlinear $\Phi_{\nu}$ would have to be corrected by non-Newtonian effects. The exact matching of the linear evolution of $\Phi_\nu$ for scales outside the horizon and the nonlinear results would require a nonlinear analysis which also includes post-Newtonian terms. This unknown region $10^{-3} \lsim k/h \lsim 6\cdot10^{-3}$ may leave the strongest observational imprint by the ISW effect.

Fig.\ref{fig:grav_pot_of_k} can be taken as a clear illustration that the formation of nonlinear structures makes the output of linear Boltzmann codes unreliable when density fluctuations become nonlinear. Instead of increasing continuosly, the gravitational potential will saturate at the typical values found by the nonlinear analysis. An unjustified extrapolation of the linear approximation overestimates the gravitational potential by many orders of magnitude for scales of a few hundred Mpc. For the parameters used in this paper, $m_\nu = 2.1$ eV, $\beta = -52$, the linear extrapolation would predict enormous oscillations in the spectrum of CMB anisotropies for angular momenta $l \approx 100$, very strongly in contradiction with observation. Placing reasonable bounds on the gravitational potential, consistent with our findings for the nonlinear evolution, completely eliminates this feature and makes the spectrum of anisotropies insensitive to growing neutrinos in this range of $l$.

On the other hand, in the range of smaller $l \approx 10$, a large time variation of the gravitational potential for scales $10^{-3} \text{ h/Mpc} \lsim k \lsim 10^{-2}$ h/Mpc could leave a big imprint on the CMB spectrum. A too large ISW effect could rule out such a scenario. We conclude that in principle the ISW effect provides a viable mean to constrain growing neutrino cosmologies, at least in the case of large neutrino masses. For smaller neutrino masses the neutrino fraction $\Omega_\nu$ of the energy density is smaller, thereby presumably reducing the ISW signal.

\section{Initial conditions} \label{sec:ini_conds}

\subsection{Matching of the radius}

The size of a given neutrino lump under investigation is determined by the parameter $r_{in}$. For a comparison with linear perturbation theory, as in fig.\ref{fig:grav_pot_of_scale}, we have to relate $r_{in}$ to an appropriate comoving wave number $k$. We will see in the next section that for a given scale there is a range of redshifts at which the fully relativistic linear and the Newtonian nonlinear evolution coincide. We can make use of this in order to approximately fix the relation between our initial width parameter $r_{in}$ and the comoving momentum scale $k$ for the associated linear fluctuation. For this purpose we choose a redshift $z_m$ within the discussed range. At the given redshift $z_m$ we choose $r_{in}$ such that the nonlinear and linear quantities relate via
\be r_{in} =  \frac{\pi}{k} \pp \ee
Note that the exact choice of $z_m$ is not important. Within the linear regime $R/a = const$, since the radius scales with the background scale factor. Hence, different matching redshifts yield equivalent results for the relation between $r_{in}$ and $k$.

\subsection{Matching of the amplitude}

\begin{figure}[ht] 
\begin{center}
\includegraphics[width=85mm,angle=-0.]{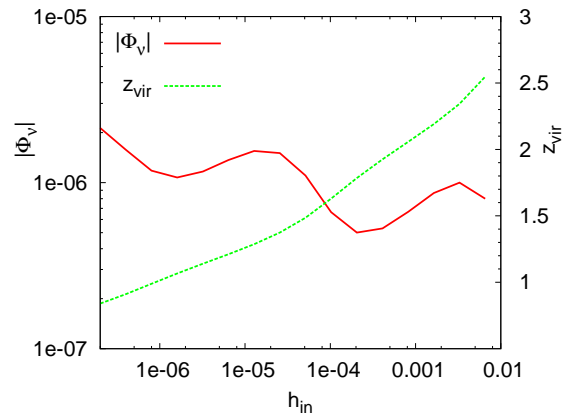}
\\
\end{center}
\caption{Dependence of the gravitational potential at virialization and virialization redshift $z_{vir}$ on the initial amplitude $h_{in}$ for $r_{in} = 45$ Mpc. The oscillations in $\Phi_{\nu}$ reflect the oscillation of the neutrino mass as a function of $z_{vir}$.}
\label{fig:universal_hin}
\vspace{0.5cm}
\end{figure}

Taking the radius of the overdensity as a free parameter, the amplitude $h_{in}$ still needs to be fixed. From a primordial gaussian fluctuation spectrum we expect a characteristic distribution of initial amplitudes around zero, with a mean square dispersion given by $\sigma_{lin}(r_{in})$. Actually, we find that the redshift of virialization $z_{vir}$ depends monotonically on $h_{in}$, while the final gravitational potential of the lump remains rather insensitive with respect to the precise choice of $h_{in}$. We demonstrate this in fig.{\ref{fig:universal_hin}}. This approximate `universality' of the lump properties after collapse may be of substantial help for observational investigations of our scenario.

\begin{figure}[ht] 
\begin{center}
\includegraphics[width=85mm,angle=-0.]{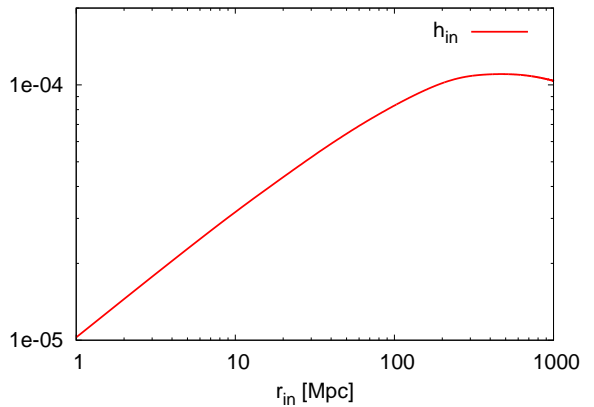}
\\
\end{center}
\caption{Relation between initial radial parameter $r_{in}$ in comoving units and dominant initial amplitude $h_{in}$.}
\label{fig:hin_of_rin}
\vspace{0.5cm}
\end{figure}

A good estimate for a characteristic amplitude $h_{in}$ can be obtained as follows. 
We consider the overall perturbation $\sigma(r_{in}, h_{in})$ defined as the convolution of our initial gaussian density configuration $\delta({\bf{x}})$ and a window function of size $r_{in}$:
\be \label{sigma_def} \sigma^2(r_{in}, h_{in}) = \frac{1}{(2 \pi)^3} \int_V{|\delta(k)|^2 |W(k)|^2 d^3k} \vv \ee
where $\delta({\bf k})$ and $W({\bf k})$ are the Fourier transforms of the density configuration $\delta({\bf{x}})$ and of the window function $W(r_{in})$ respectively. The best estimate for $h_{in}$ corresponds to the value for which $\sigma(r_{in}, h_{in}) = \sigma_{lin}(r_{in})$. Here $\sigma_{lin}(r_{in})$ is computed by using in eq.(\ref{sigma_def}) the same window function $W(r_{in})$, but now the linear fluctuation spectrum $\delta_{lin}(k)$ instead of the Fourier transform of eq.(\ref{eq:dens_ini}). The linear $\delta_{lin}(k)$ are evaluated from a Boltzmann code for growing neutrinos and taken at redshift $z_{in}$.
Note that the matching is done in $k$-space and therefore requires no explicit calculation of the relation between length scales $r$ and momentum $k$. Since $\sigma^2(r_{in},h_{in})\sim h_{in}^2$ we obtain a simple estimate for a characteristic initial amplitude. The dependence of the initial amplitude $h_{in}$ on the initial radius $r_{in}$ is displayed in fig.\ref{fig:hin_of_rin}.

Our choice of $h_{in}$ neglects two effects: the dragging of neutrinos by dark matter fluctuations and the nonzero pressure at early stages of the evolution. We have checked that these effects largely cancel. This is demonstrated by the good agreement with full linear perturbation theory (including both effects) in the matching range, as discussed in the next section.

In concluding this section, we mention a series of checks that have been done in order to verify the stability of our results.
\begin{itemize}
\item We have checked how the formation of the lump depends on different initial velocities. In particular, increasing initial nonradial velocities leads to a faster collapse. All features, such as size and total mass of the final lump remain unaltered. In particular, the final gravitational potential is unaffected by a change of initial velocities.
\item We have investigated the formation of two lumps within our box, both in the case in which they are far enough apart to evolve independently and in the case in which they are close enough to merge. For a given final size $R_f$ of the lump, neutrinos distribute equivalently in the two cases, with no relevant effect on the gravitational potential. This scenario does not cover, however, the interesting and probably realistic case of many lumps moving for a while under their mutually attractive cosmon forces.
\item We have considered the case of an initially nonspherical profile, with a pronounced ellipticity as in the following expression \be \label{eq:dens_ell} \delta_{\nu}(\mathbf{x}) = h_{\nu, in}e^{-\frac{x^2+y^2}{w_{xy}^2}-\frac{z^2}{w_z^2}} \vv \nonumber \ee with $w_z \gg w_{xy}$. Again, for a given $R_f$ the results do not seem to be affected in an important way.
\item Finally, we have estimated the effect of Fermi pressure within a Thomas-Fermi approximation as a function of the lump size. The ratio of Fermi force and fifth force, increasing with the scale, never happens to be bigger than a few percent and therefore has a negligible effect on the estimated gravitational potential. 
\end{itemize}

\section{Relativistic equations} \label{sec:rel_nonrel}
In this section we further discuss the range of validity of the numerical integration and clarify the choice of initial conditions. We are interested in the range of scales and redshifts where neutrino perturbations are expected to grow nonlinear as found in \cite{mota_etal_2008}, when neutrinos are nonrelativistic and pressure terms are negligible. 
The integration of the nonlinear equations starts inside the linear regime, when neutrinos start becoming nonrelativistic, at $z_{in} \sim 9$.
At this time the system is still correctly described by the set of fully relativistic linear perturbation equations, in which neutrino pressure terms are included. We want to make sure to have a proper matching of the regime where we can trust the nonlinear equations used in the numerical solution to the regime where the relativistic linear equations are valid.

For fast comparison to the nonlinear equations we recall here the set of linear perturbation equations specified in the case in which the Newtonian limit applies. (For the full, closed, system of linear perturbation equations see \cite{mota_etal_2008}, where linear perturbations for growing neutrinos have been extensively investigated.) They read 
\bea \delta_{\phi}' &=&  3 {\cal H} (w_{\phi} - c_{{{\phi}}}^2) \delta_{\phi} -(1+ w_\phi) k v_{\phi} \\ &-& \beta \phi' \frac{\rho_\nu}{\rho_{\phi}} \left[(1-3 w_\nu) \delta_{\phi} - (1-3 c_{{\nu}}^2) \delta_\nu \right]  \nonumber  \vv \label{eq:lin_deltaphi} \eea
\bea
\delta_{\nu}' = 3 \left({\cal H} - \beta \phi'\right) (w_{\nu} - c_{{\nu}}^2) \delta_{\nu} - (1+w_{\nu}) k v_\nu  \label{eq:lin_delta} \pp
\eea
In deriving these equations, we have defined the line element as given by $ds^2 = -a^2[(1 + 2 \Psi) d\tau^2 - (1 + 2 \Phi)\gamma_{ij} dx^i dx^j]$, where $\Phi$ corresponds to the usual gravitational potential and we are working in Newtonian gauge. The equations for the density contrasts $\delta_i(k) = \frac{1}{V}\int{\delta_i({\bf x}) exp(-i {\bf{k \cdot x}}) d^3x }$ (defined as the Fourier transform of the local density perturbation $\delta_i({\bf x}) = \delta \rho_i(x)/\rho_i(x)$ over a volume $V$) involve the velocity perturbations, which evolve according to
\bea 
v_{\phi}' &=& -{\cal H}(1-3 w_{\phi}) v_{\phi} - \beta \phi'(1-3 w_\nu) \frac{\rho_\nu}{\rho_{\phi}} v_{\phi}  \nonumber \\ &-&  \frac{w_{\phi}'}{1+w_{\phi}}v_{\phi}  + k c_{{{\phi}}}^2 \frac{\delta_{\phi}}{1+w_{\phi}} +  k {\bf \Psi}  \nonumber \\ &-&  \frac{2}{3} \frac{w_{\phi}}{1+w_{\phi}} k \pi_{T_{\phi}} +  \beta k \delta \phi \frac{\rho_\nu}{\rho_{\phi}} \frac{1-3 w_\nu}{1+w_{\phi}}  \vv
\\ 
v_{\nu}'  &=& (1 - 3 w_\nu) (\beta \phi' - {\cal H}) v_\nu -  \frac{w_{\nu}'}{1+w_{\nu}}v_\nu   \nonumber \\ &+& k c_{{\nu}}^2 \frac{\delta_{\nu}}{1+w_{\nu}} + k {\bf{\Psi}} - \frac{2}{3} k \frac{w_{\nu}}{1+w_{\nu}} \pi_{T \, \nu} \nonumber  \\ \label{eq:lin_v} &-& \beta k \delta \phi \frac{1-3 w_\nu}{1+w_\nu} \pp \eea

These equations reduce to the linearized Navier Stokes equations if we set neutrino pressure terms to zero and consider the case of no anisotropic stress for which $\Psi = -\Phi$, thus obtaining for the neutrino component:
\bea 
\delta_{\nu}' &=&  - k v_\nu  \label{eq:lin_delta_newt} \vv
\\
v_{\nu}'  &=& -({\cal H}- \beta \phi') v_\nu  - k ({\bf{\Phi}} + \beta \delta \phi) \label{eq:lin_v_newt} \pp 
\eea
The latter equations, here written in Fourier space, are equal to the ones obtained by linearizing the Navier Stokes equations (\ref{eq:com_ns1}, \ref{eq:com_ns2}). 

For a certain first time period after we start the integration of the nonlinear equations, neutrino pressure terms will not yet be negligible and therefore the output of the linear equations (\ref{eq:lin_delta},\ref{eq:lin_v}) will differ from the output of the nonlinear equations (\ref{eq:com_ns1}-\ref{eq:com_grav_pot}), in which we do not include pressure terms. We will therefore consider the nonlinear equations without pressure terms as reliable only after a certain redshift $z_{nl}$, while prior to $z_{l}$ we trust the relativistic linear equations. There is a redshift range where both equations are valid. We match the evolution at a certain redshift $z_m$ in the range $z_l < z_m < z_{nl}$, which depends on the size of the lump under investigation.
This matching is done by choosing appropriate initial conditions for the numerical solution. More precisely, we require for the `matching' redshift $z_m(k)$ that pressure terms effectively become negligible. For $z<z_m$ we will then consider the output of the nonlinear integration as valid.

\begin{center}
\begin{figure}[ht]
\begin{picture}(0,200)(140,0)
\includegraphics[height=85mm, angle=90]{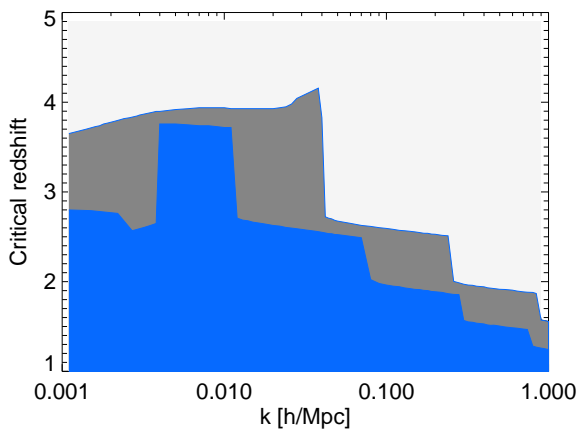}
\end{picture}
\\
\caption{Critical redshift versus momentum scale. The critical redshift is defined as the $z$ at which pressure terms become negligible in the equation for $v_\nu$, for each scale. The blue (dark shaded) region corresponds to relative pressure contributions less than $1\%$; the grey (intermediate shaded) region corresponds to relative pressure contributions between $1\%$ and $10\%$. The above light grey region corresponds to pressure contributions $> 10\%$.}
\label{fig_deltacrit}
\vspace{0.5cm}
\end{figure} 
\end{center}

In fig.\ref{fig_deltacrit} we show the critical redshift $z_\text{cr}$ for which neutrino fluctuations become well approximated by pressureless nonrelativistic particles. The critical redshift is plotted as a function of the length scale defined as $R/a \equiv \pi/k$, where $k$ is the momentum scale. We define the critical redshift as the value of $z$ at which the relative size of pressure terms drops below a certain value in the linear equation for $v_\nu$ (\ref{eq:lin_v}). The shaded region corresponds to relative pressure contributions between $1\%$ and $10\%$. Above the upper (green) line pressure contributions are $> 10\%$; below the bottom (blue) line pressure contributes for less than $1\%$.

Our nonlinear equations are meaningful for those scales and redshifts at which radiation, baryons and CDM density perturbations are subdominant with respect to neutrino density perturbations, as it happens for large scales at late times due to the rapid effect of the fifth force acting on neutrinos only. Only in this regime it is justified to restrict the contributions to the gravitational potential in (\ref{eq:com_grav_pot}) to neutrinos. 
Finally, note that although lumps are expected to form at very large scales ($\sim 200$ Mpc), these are still well within the horizon, which justifies a Newtonian approach.

In closing this section, we would like to mention that it is also possible to show that the nonlinear equations described by (\ref{eq:com_ns1} - \ref{eq:com_grav_pot}) follow from the set of full relativistic equations taken in the Newtonian limit. We do not intend to show this explicitly here, but just outline the basic procedure.
For the derivation of the nonlinear relativistic equations we consider the case in which there is no anisotropic stress, $\Psi = -\Phi$ so that the line element takes the form $ds^2 = -a^2[(1 - 2 \Phi) d\tau^2 - (1+2 \Phi)\gamma_{ij} dx^i dx^j]$. 
We consider the equations in the Newtonian limit, restricting our analysis to spatial scales much smaller than the horizon size, that is to say when ${\cal H}/k \ll 1$.
Furthermore, we consider the weak field approximation, in which the scalar fluctuation $\delta\phi$ and the gravitational potential $\Phi_{\nu}$ are considered to be small quantities and only enter the equations up to linear order. We consider perturbation terms up to second order in $\mathbf{v_{\nu}}(\tau,\mathbf{x})$, as in the Newtonian limit velocities are small with respect to light speed (i.e. $\ll 1$ according to our convention).
The stress energy tensor of the coupled neutrino fluid,  $T_{(\nu)\,\alpha}^{\,\,\,\,\,\,\,\beta}$, then takes on the following form

\bea
T_{(\nu)\,0}^{\,\,\,\,\,\,\,0} &=& -\bar\rho_{\nu}\left(1 + \delta_{\nu}\right) - {\cal B}(\bar{\rho}_\nu, \bar{p}_\nu, \delta_\nu) \mathbf{v_{\nu}}^2 \\
T_{(\nu)\,0}^{\,\,\,\,\,\,\,i} &=& - {\cal B}(\bar{\rho}_\nu, \bar{p}_\nu, \delta_\nu) (\mathrm{v}_{\nu})^i \nn\\
T_{(\nu)\,i}^{\,\,\,\,\,\,\,0} &=& \left(1 - 4\Phi\right) {\cal B}(\bar{\rho}_\nu, \bar{p}_\nu, \delta_\nu) (\mathrm{v}_{\nu})_i \nn\\
T_{(\nu)\,i}^{\,\,\,\,\,\,\,j} &=& \left(\bar{p}_{\nu} + \delta{p}_{\nu}\right)\delta_i^j
\label{eq:T_nu} + {\cal B}(\bar{\rho}_\nu, \bar{p}_\nu, \delta_\nu) (\mathrm{v}_{\nu})_i\,(\mathrm{v}_{\nu})^j \nn
\eea
where we have defined  \be {\cal B}(\bar{\rho}_\nu, \bar{p}_\nu, \delta_\nu) \equiv \left(\bar\rho_{\nu} + \bar{p}_{\nu} + \bar\rho_{\nu} \delta_{\nu} (1 + c_s^2)\right) \ee and in the linear regime $c_s^2 = \delta\rho_\nu/\delta{p}_\nu$ corresponds to the squared sound velocity of neutrino perturbations.


In the Appendix, perturbation equations are derived from the conservation equation (\ref{eq:ps_cons}), $\nabla_{\gamma}T_{\alpha}^{\gamma} = -\frac{\beta}{M}  T_{\gamma}^{\gamma}  \partial_{\alpha}\phi$. 
When pressure terms are neglected and for $\delta \rho_c \ll \delta \rho_\nu $, these equations reduce to the Navier Stokes set of equations (\ref{eq:com_ns1} - \ref{eq:com_grav_pot}), in which the dragging term $\beta\phi'\mathbf{v}_{\nu}$ explicitly appears. The latter was found to be a small effect for the chosen values of the coupling, since within the considered redshift range the fast oscillations of the scalar field average out and, as discussed before, $\phi$ can roughly be considered to be constant.

\section{Conclusions} \label{conclusions}
In growing neutrino cosmologies, the neutrino mass grows in time as a function of a light dark energy scalar field, the cosmon. The resulting coupling between cosmon and neutrinos modifies the evolution of the cosmon as soon as neutrinos become nonrelativistic. From this time on the cosmon evolves only very slowly, and its potential energy acts similar to a cosmological constant. This can provide a natural explanation of the coincidence problem without effectively introducing new parameters. The current dark energy density can be related to the neutrino mass. In addition, the interaction gives rise to a new long-range attractive force for neutrinos. This fifth force is responsible for a rapid growth of neutrino perturbations. These perturbations become nonlinear on large length scales and may eventually form very large stable structures, neutrino lumps.

We have extended previous studies on the topic to the nonlinear regime, providing a method to investigate the formation of neutrino lumps by evaluating Navier Stokes equations, in which the dragging term due to the coupling has been suitably included. As we have verified, these are the most general equations to compute the spacetime evolution of perturbations subject to external forces within the Newtonian limit. Unlike spherical collapse methods, the full nonlinear equations describe coherently the growth of structures due to an external force stronger than gravity. We note that although our calculations were performed within a growing neutrino scenario, the results are not limited to this case. Instead, the method can be applied for general coupled quintessence models, or whenever a force different from gravity is present, driving the growth of nonrelativistic matter perturbations. 

We have numerically solved the set of hydrodynamical equations on a three-dimensional spatial grid. This has allowed us to follow the evolution of a single neutrino lump in physical space and to determine its properties as it approaches stability. Nonrelativistic neutrinos decouple from the background expansion and collapse into virialized stable structures. We have identified the characteristic gravitational potential of these structures as a function of redshift and lump size and we have illustrated the detailed behavior of the lumps from the linear regime to virialization. Indeed, after a period of fifth-force driven collapse, the build-up of large nonradial velocities leads to stabilization of the lumps. By solving on a three-dimensional grid with a general initial profile, we are able to provide a detailed illustration of the evolution at different distances from the center of the lump, following the change in kinetic and potential energy as well as the increasing contribution of nonradial velocities. We have also estimated the density profile of the virialized neutrino lumps and the associated profile of the gravitational potential. Limitations of our results for characteristic properties of simple neutrino lumps arise from two issues. First, at the present stage the numerical precision of our algorithm does not allow us to compute the behavior after virialization sets in and to follow the evolution until the neutrino lump becomes approximately stable. Second, our work concentrates on the evolution of single lumps while we have not addressed interesting topics on cosmological scales such as the dynamics of the lumps and their possible merging.

The cosmological gravitational potential provides an important mean to test the model versus observations and can be used to estimate the impact of neutrino lumps on the CMB angular power spectrum. An extrapolation from the gravitational potential for single lumps to the neutrino lump induced cosmological potential $\Phi_{\nu}(k)$ at a given wave number $k$ involves an averaging over the distribution of neutrino lumps. The present work can only give a very rough estimate of the result of this averaging. Nevertheless, it is apparent that the values of $\Phi_{\nu}(k)$ resulting from the nonlinear hydrodynamical equations remain many orders of magnitude below the extrapolation of the linearized equations if $k \gg 10^{-3}$ Mpc$^{-1}$. We find a typical value $\Phi_{\nu}(k) = 10^{-6} (k/k_0)^{-2}$, where the large uncertainty is reflected by the uncertainty in $k_0$ that we estimate in the range $8.6\cdot10^{-3}\text{ Mpc}^{-1} < k_0 < 5.6\cdot10^{-2}\text{ Mpc}^{-1}$. This estimate has to be modified for $k \lsim 4.5\cdot10^{-3}$ Mpc$^{-1}$, where neglected effects beyond the Newtonian approximation become relevant, and for $k \gsim 1.1$ Mpc$^{-1}$, where neutrino free streaming prevents the formation of lumps. Also, a formation history involving merging may reduce $\Phi_{\nu}(k)$, especially for small $k$ corresponding to very large lumps.

The characteristic size of $\Phi_{\nu}(k)$ found in this paper is of an order of magnitude where it may leave imprints on the CMB spectrum at small angular momenta. In particular, the late ISW effect could be enhanced, leading to stronger correlations between temperature fluctuations in the CMB and observed large scale structures in the gravitational potential. (The present observational situation hints to an enhancement of this correlation by a factor of about $2$ as compared to the $\Lambda$CDM model \cite{ho_etal_2008}.) Small oscillations of the CMB spectrum for angular momenta $l \lsim 50$ are also conceivable. Without a more reliable estimate of $\Phi_{\nu}(k,z)$ it seems premature to judge if a given growing neutrino model remains compatible with observation. We recall in this context that the present work concentrates on a particular class of growing neutrino models (constant $\beta$) and assumes a large average present neutrino mass $\bar{m}_{\nu}(t_0) \sim 2$ eV. We expect that a smaller $m_\nu$ reduces the cosmological effects of neutrino lumps since the total neutrino fraction of the energy density $\Omega_\nu$ gets reduced and the neutrino induced effects set in at even more recent cosmological times.

It may be possible to directly observe large neutrino lumps through their gravitational potential. A possible indication would be an observation of structures at large length scales where fluctuations in the $\Lambda$CDM model are expected to remain linear such that substantial over- or underdensities are very rare. The gravitational potential of large neutrino lumps may influence a substantial fraction of space. If we are sitting not too far from a neutrino lump (or even within a neutrino lump) this may induce a certain anisotropy of the observed sky on the largest scales (i.e. small differences between northern and southern hemisphere or similar effects). Finally, the rapid recent growth of the neutrino induced gravitational potential could lead to an enhancement of the peculiar velocities as a characteristic effect of the nongravitational interactions \cite{ayaita_etal_2009}. Present observations of the large scale bulk flow seem to suggest values substantially larger than expected in the $\Lambda$CDM model \cite{perivolaropoulos_2008,watkins_2008,kashlinsky_2008,lavaux_2008}. Finally, the cosmon may also have a coupling to dark matter, substantially smaller than the cosmon-neutrino coupling. In this case the fifth force effects could also influence the behavior of dark matter. Detection of such effects would challenge standard $\Lambda$CDM models, providing a hint for interacting cosmologies such as growing neutrino models. 

Let us end with the remark that there is a chance that the time variation of the neutrino mass could even be detected. Indeed, our understanding of structure formation and other cosmological features places strong upper bounds on the neutrino masses in early cosmology, say for $z \gtrsim 5$. One may argue about the precise location of this bound, but an average neutrino mass $m_{\nu}(z > 5)$ of $0.5$ eV would certainly have left a strong imprint on cosmology which has not been observed. In consequence, if the direct searches for a neutrino mass or the neutrinoless double beta decay indicate a present neutrino mass larger than $0.5$ eV, this could be interpreted as a strong signal in favor of a growing neutrino mass.
\\
\\
\\

\begin{acknowledgments}
VP acknowledges the Alexander von Humboldt Foundation.
We thank Lily Schrempp for precious contributions during the first stages of this work.
\end{acknowledgments}

\section*{APPENDIX}

\subsection*{Nonlinear relativistic perturbation equations}
The evolution equations for a neutrino overdensity can be derived from conservation of the energy-momentum tensor (\ref{eq:T_nu}):
\begin{widetext}
\bea
\delta_{\nu}' &=& 3 \left({\cal H}-\beta \phi'\right) \left(w_{\nu}-c_s^2\right) \delta_{\nu} - (1 + w_{\nu})(1 + R_{\nu}\delta_{\nu})\left(\nabla\mathbf{v_{\nu}}-3 \Phi '\right) - \beta \delta\phi' (1 - 3w_{\nu} + \delta_{\nu}(1 - 3 c_s^2)) \nn \\
&+& \mathbf{v}_{\nu}\left[-\nabla\delta_{\nu} + (1 + 8\Phi)\frac{\nabla\delta{p}_{\nu}}{\rho_{\nu}} + 4(1+w_{\nu})(1+R_{\nu}\delta_{\nu})\nabla\Phi - 2\beta\,\nabla\delta\phi\, (1 - 3w_{\nu} + \delta_{\nu}(1 - 3 c_s^2))\right] \nn \\
\label{eq:nl_delta} &+& \mathbf{v}_{\nu}^2\left[\delta_{\nu} c_s^{2'} + w_{\nu}' + (1-3w_{\nu})(1+w_{\nu})(1+R_{\nu}\delta_{\nu})({\cal H} - \beta\phi') + 3\delta_{\nu}(1+c_s^2)(w_{\nu}-c_s^2)({\cal H}-\beta\phi')\right] \vv \eea
\bea
\mathbf{v_{\nu}}' &=& -\nabla\Phi + \frac{1 - 3 w_{\nu} + \delta_{\nu}(1 - 3 c_s^2)}{(1+w_{\nu})(1+R_{\nu}\delta_{\nu})}\beta\nabla\delta\phi - \frac{1 + 4\Phi}{\rho_{\nu}(1+w_{\nu})(1+R_{\nu}\delta_{\nu})}\nabla\delta{p}_{\nu} \nn \\
&-& \left[\scalprod{v_{\nu}}{\nabla} + \frac{(2+c_s^2)\mathbf{v}_{\nu}\nabla\delta{p}_{\nu} - \rho_{\nu}\,c_s^2\mathbf{v}_{\nu}\nabla\delta_{\nu}}{\rho_{\nu}(1+w_{\nu})(1+R_{\nu}\delta_{\nu})} - 3(1 - c_s^2)\Phi' + ({\cal H} - \beta\,\phi')\,\frac{1 - 3w_{\nu} + R_{\nu}\,\delta_{\nu}\,(1 - 3 c_s^2)}{1 + R_{\nu}\delta_{\nu}}\right]\,\mathbf{v}_{\nu} \nn \\
\label{eq:nl_v} &+& \left[c_s^2\divvec{v_{\nu}} - \frac{\delta_{\nu}c_s^{2'} + w_\nu'}{(1+w_{\nu})(1 + R_{\nu}\delta_{\nu})} + \frac{R_{\nu}(1 - 3w_{\nu} + \delta_{\nu}(1 - 3 c_s^2))}{1 + R_{\nu}\delta_{\nu}}\,\beta\,\delta\phi'\right]\mathbf{v}_{\nu} \vv
\eea
\end{widetext}
where we introduced $R_{\nu} \equiv (1 + c_s^2)/(1 + w_{\nu})$. The quantity $c_s^2 = \delta{p_{\nu}}/\delta\rho_{\nu}$ equals the squared sound speed of neutrino perturbations on a linear level.

One may verify that (\ref{eq:nl_delta}) and (\ref{eq:nl_v}), if decomposed into Fourier modes, reduce to (\ref{eq:lin_delta}) and (\ref{eq:lin_v}) in the linear regime.

The 0-0-component of Einstein's field equations fulfills
\be
\Delta \Phi = \frac{a^2}{2}\delta\rho + \frac{a^2}{2} \rho  \left(\mathbf{v}^2+2 (1 + \delta) \Phi \right) + 3 {\cal H} \Phi'
\ee
while the perturbed Klein Gordon equation yields
\begin{widetext}
\be
\delta\phi'' = a^2 \beta \rho_{\nu} (\delta_{\nu}(1 - 3 c_s^2) + 2 (1-3 w_{\nu})\Phi) - 2 a^2 \Phi U_{,\phi } - a^2 \delta\phi U_{,\phi\phi} + \Delta \delta\phi - 2 {\cal H}\delta\phi'+ 4 \phi' \Phi' \pp
\ee
\end{widetext}
If pressure and sound speed are negligible with respect to the respective densities, we may omit all terms proportional to $p_{\nu} = w_{\nu}\rho_{\nu}$, $\delta{p_{\nu}} = c_s^2\,\delta\rho_{\nu}$ and their derivatives. Subsequently considering the Newtonian limit yields eqs. (\ref{eq:com_ns1})-(\ref{eq:com_grav_pot}).

\end{document}